\documentclass[twocolumn]{article}

\usepackage{graphicx}

\def\la
  {\mathrel{\hbox{\rlap{\hbox{\lower4pt\hbox{$\sim$}}}\hbox{$<$}}}}
\def\ga
  {\mathrel{\hbox{\rlap{\hbox{\lower5pt\hbox{$\sim$}}}\hbox{$>$}}}}

\title
{Supernova Ia:\\ a Converging Delayed Detonation Wave}

\author{ N.V.Dunina-Barkovskaya,
V.S.Imshennik\thanks{email: imshennik\symbol{64}vxitep.itep.ru},
S.I.Blinnikov %\thanks{email: blinn\symbol{64}sai.msu.su}
\\
 {\small \it Institute for Theoretical and
Experimental Physics, Moscow, Russia}\\ {\small \it
Max-Planck-Institut f\"ur Astrophysik, Garching, Germany}\\
{\footnotesize Submitted to Astronomy Letters on 25.12.2000}}
\date{}
\begin{document}
\maketitle

\begin{abstract}

A model of a carbon-oxygen (C--O) presupernova core with an
initial mass $1.33 M_\odot$, an initial carbon mass fraction
$X^{(0)}_{\rm C}=0.27$, and with an average mass growth-rate
  $5\cdot 10^{-7} M_\odot\mbox{ yr}^{-1}$
  due to accretion in a binary system
was evolved from initial
 density
 $\rho_{\rm c}=10^9\mbox{g cm}^{-3}$,
 and temperature $T_{\rm c}=2.05\cdot 10^8$K
through convective
  core formation and its subsequent expansion to the carbon runaway at
 the center.
The only thermonuclear reaction contained in the equations of
evolution
 and runaway was the carbon burning reaction $^{12}$C$+^{12}$C
with an energy release corresponding to the full transition of
carbon and oxygen (with the same rate as carbon) into $^{56}$Ni.
As a parameter we take  $\alpha_{\rm c}$ --- a ratio of a mixing
length to the size of the convective zone. In spite of the crude
assumptions, we obtained a pattern of the runaway acceptable for
the supernova theory with the strong dependence of its duration on
$\alpha_{\rm c}$ . In the variants with large enough values of
$\alpha_{\rm c}=4.0 \cdot 10^{-3}$  and $3.0 \cdot 10^{-3}$ the
fuel combustion occurred from the very beginning as a prompt
detonation. In the range of
 $2.0\cdot 10^{-3} \ge \alpha_{\rm c} \ge 3.0\cdot 10^{-4}$
the burning started as a deflagration with excitation of stellar
pulsations with growing amplitude. Eventually, the detonation set
in, which was activated near the surface layers of the
presupernova
 (with $m \simeq 1.33 M_\odot$ )
and penetrated into the star
 down to the deflagration front.
Excitation of model pulsations and formation of a detonation
 front are described in detail for the variant with
$\alpha_{\rm c}=1.0\cdot 10^{-3}$.
\end{abstract}

{\it Keywords:} supernovae and supernova remnants; plasma
astrophysics; hydrodynamics and shock waves; detonation and
deflagration.

\section*{Introduction}

Evolution of a degenerate C--O stellar core with a mass close to
the Chandrasekhar limit ($1.43 M_\odot$ for a carbon white dwarf,
see Bisnovatyi-Kogan, 1989) leads to the development of thermal
instability  and to the explosion. The critical density
$\rho_{\rm c \; cr}$ for the beginning of the explosion  %sb
may vary from $\sim 2\cdot 10^9$ to $\sim 10^{10} \mbox{ g
cm}^{-3}$. It increases with reduction of $\dot M$ --- the core
mass growth-rate which for a single asymptotic giant branch
 (AGB) star is defined by the Paczy\'nski --- Uus relation
(Paczy\'nski, 1970)

\begin{equation}
\label{pacuus} \dot M = 6\cdot 10^{-7} (M/M_\odot -0.5 ) M_\odot
\mbox{ yr}^{-1}\;,
\end{equation}
and  can take values from $10^{-8}$ to  $5\cdot 10^{-7}
M_\odot\mbox{ yr}^{-1}$ in an accreting white dwarf (Iben, 1982;
Hachisu et al., 1996). In the present paper we assume $\dot M =
5\cdot 10^{-7} M_\odot\mbox{ yr}^{-1}$ which can be appropriate
both for a single AGB-star and for a component of a binary system.

The explosion calculations, some of which accounted for
convection,
 were carried by various authors (Arnett, 1969; Ivanova et al., 1974;
 Nomoto et al., 1976;
 see also the review by Niemeyer and Woosley (1997)).
Ivanova et al.(1974) found that carbon was burning in the
deflagration regime with excitation of pulsations, but convection
at the supernova stage was not considered, and the initial
temperature profile was obtained not from evolutionary
calculations, but from an estimate of a convection contribution.

Woosley (1997) has carried out a series of one-dimensional
explosion calculations for accreting white dwarfs (unfortunately,
in this work the accretion rates are not specified)
 with
critical central densities from  $2\cdot 10^9$ to $8.2\cdot 10^9
\mbox{g cm}^{-3}$ and with nuclear reactions for 442 isotopes, but
he considered only the adiabatic convection at the presupernova
stage, while at the beginning of the thermal runaway
 (when $T_{\rm c}=7\cdot 10^8$ K) he turned off the convection
artificially (because further maintenance  of the adiabatic
temperature gradient in his model led to the prompt detonation),
and defined the burning front propagation velocity according to Woosley and
Timmes (1992) with a provision  for a fractal dimension of the front.

\section*{Main equations of the convective hydrodynamic model}

Many authors (e.g. Iben, 1982)  use the hydrostatic system of
equations with adiabatic convection in the stellar core, when
calculating the presupernova evolution (with time-scale $\sim
10^4$ yr) which is valid for this stage, but is too crude at the
beginning of the thermal runaway. We carried our calculations both
on the presupernova stage and on the stage of explosive burning by
the hydrodynamic scheme elaborated in papers by Blinnikov and
Rudzsky (1984), Blinnikov and Bartunov (1993), and by Blinnikov
and Dunina-Barkovskaya (1993, 1994). In the latter two papers
 this hydrodynamic scheme was used for the calculation of white
 dwarf evolution.
Here we have included non-adiabatic time-dependent
convection in well-known
mixing length approximation:
\begin{eqnarray}
\partial r/\partial t &=&v\;, \label{rdot} \\
\partial v/\partial t &=& -Gm/r^2 - 4\pi r^2(\partial P/\partial m)\;,
\label{vdot}\\
\partial T/\partial t &=& \left[ \varepsilon_{\rm CC}
                       +\varepsilon_\nu +\varepsilon_{\rm g}
                       -4\pi \partial(r^2 F_{\rm conv})/\partial m
                          \right. \nonumber \\
                      &-& \left.
                       4\pi \partial(r^2 F_{\rm rad})/\partial m
                       \right]
                       /(\partial E/\partial T)_\rho\,,
\label{tdot}\\
\partial X_{\rm C}/\partial t &=& - X_{\rm C}^2 r_{\rm CC}
+(\partial X_{\rm C}/\partial t)_{\rm conv}\;,
\label{xcdot}\\
\partial u_{\rm c}/\partial t &=& 2(v_{\rm c}^2-u_{\rm c}^2)/l_{\rm mix}\;,
\label{ucdot}
\end{eqnarray}
where $X_{\rm C}$ is the mass fraction of $^{12}{\rm C}$,
$\varepsilon_{\rm CC}$ is the energy release during the carbon
burning, $\varepsilon_\nu$ stands for standard neutrino energy
losses (as approximated by Schinder et al., 1987; see also Haft et
al., 1994), $\varepsilon_{\rm g}$ is the energy liberation during
the gravitational contraction, $F_{\rm rad}$ is the radiative
energy flux, $F_{\rm conv}$ is the convective energy flux, $u_{\rm
c}$ is the time-dependent convective velocity (see Arnett 1969),
and $v_{\rm c}$ is the stationary convective velocity in the
mixing length approximation which is given by the formula (see,
e.g.,
 Bisnovatyi-Kogan, 1989)
\begin{equation}
\label{vconv} v_{\rm c}=(g(\partial \ln \rho/\partial r
          - (\partial \ln \rho/\partial r)_S))^{1/2}
           \cdot l_{\rm mix}/2\;.
\end{equation}

The convective energy flux $F_{\rm conv}$ and the value $(\partial
X_{\rm C}/\partial t)_{\rm conv}$, which corresponds to the change
of a carbon mass fraction $X_{\rm C}$ by convection, were
calculated by the formulas
\begin{equation}
\label{fconv}
 F_{\rm conv} = ((\partial T/\partial r)_S - \partial T/\partial r)
              c_p \rho u_{\rm c} l_{\rm mix}/2\;,
\end{equation}

\begin{equation}
\label{dxcconv}
\left(\frac{\partial X_{\rm C}}{\partial t}\right)_{\rm conv} =
\frac{1}{r^2}\frac{\partial}{\partial r}\left(r^2 l_{\rm mix}u_{\rm c}
\frac{\partial X_{\rm C}}{\partial r}\right)\;.
\end{equation}
The arrangement and the sizes of the convective zones are defined
by the Schwarzschild criterion (according to the aforecited
formulas (\ref{vconv}) and (\ref{fconv})). On the presupernova
stage we did not account for time-dependent convection and the
formulas (\ref{fconv}--\ref{dxcconv}) contained $v_{\rm c}$
instead of $u_{\rm c}$. A comparison of calculations with
stationary and time-dependent convection on the supernova stage
was performed in (Dunina-Barkovskaya, Imshennik, 2000).

The mixing length in the $i$-th mass zone was determined
 by the relation
\begin{equation}
\label{lmixdri} l_{\rm mix}^{(i)}= {\rm min}
\left(\alpha_{\rm P}(  \partial |\ln P|/  \partial r)^{-1}_i,
\alpha_{\rm c} \Delta r_{\rm c} \right)\;,
\end{equation}
where $\Delta r_{\rm c}$ is the size of the convective zone containing
$i$-th zone, and $\alpha_{\rm P}$ in our calculations was taken equal
to unity.

Thermodynamic functions for the electron-positron gas
(the pressure $P$, the entropy $S$ etc.)
were
calculated with help of  Nadyozhin's asymptotic realtions
described by Blinnikov et al. (1996), and for the ionic gas with screening
--- by Yakovlev
and Shalybkov (1988).

\section*{Initial and boundary conditions. Computations on the
presupernova stage}

The initial state of the presupernova C--O core should to be taken
from reliable evolutionary calculations. We begin our computations
with the central density $10^9\mbox{g cm}^{-3}$ which, according
to Iben (1982), corresponds to the central temperature about
$T_{\rm c}^{(0)} = 2.05\cdot 10^8$~K. Everywhere below we define,
as a beginning of the runaway and, thereafter, as an end of the
presupernova stage, the moment of time when the temperature in the
central zone reaches the value of $5\cdot 10^9 {\rm K}$.

The initial carbon mass fraction  $X_{\rm C}^{(0)} = 0.27$
(constant throughout the C--O core) was taken the same as in
Iben's (1982) work. Assuming the above-mentioned central
parameters for the C--O core and its total mass $1.33 M_\odot$, we
have computed the adiabatic configuration in hydrostatic
equilibrium which was used as an initial condition. It should be
mentioned that its difference from Iben's (1982) model seems
insignificant.

During the calculations of the evolution of  this model with
account for a constant mass growth-rate $5\cdot 10^{-7}
M_\odot\mbox{ yr}^{-1}$ (typical for accreting white dwarfs in
binary systems), it is worth-while to set a non-zero outer
boundary condition for the pressure. In our previous work
(Dunina-Barkovskaya, Imshennik, 2000), where the calculations led
to relatively small changes of the total mass before the runaway,
we included all the accreting mass in the outer boundary
condition.

This condition is easily derived in the approximation of a thin
 ($\Delta R/R \ll 1$) and light ($\Delta M/M \ll 1$) envelope.
Really, by integrating the hydrostatic equilibrium equation
$ - {\rm d}P/{\rm d}m = Gm/4\pi r^4$ over the envelope thickness
we obtain for the pressure $P_{\rm b}$
on the outer boundary of the zone with the Lagrangean coordinate
$M$:
\begin{eqnarray}
  P_{\rm b}(M)&=&\frac{G}{4\pi}\int^{M+\Delta M}_M
\frac{m{\rm d}m}{r^4} \simeq  \frac{G}{8\pi
R^4}\left(m^2\right)^{M+\Delta M}_M \nonumber \\&\simeq &
\frac{GM}{4\pi R^4}\Delta M ,\label{pex}
\end{eqnarray}
where $\Delta M = \dot M t$ and $ R=R(t)$ during the hydrostatic
evolution, i.e. the outer pressure $P_{\rm b}$ from (\ref{pex})
increases in proportion to the time $t$ which is measured from the
beginning of the C--O core evolution calculations.

In this case the full mass $M_N + \Delta M$
(where $M_N$ is the mass of $N$ Lagrangean zones contained
in the hydrodynamic model) grew
up to  $\sim 1.351
M_\odot$, in accordance with evolution time $4.25\cdot 10^4$ yr
(Dunina-Barkovskaya, Imshennik, 2000).

In the present work, we have gradually, according to the
accretion, added new Lagrangean zones to the model (cf. Woosley,
1997). The mass of the zones decreased geometrically from $7.83
\cdot 10^{-3} M_\odot$ (151-th zone) down to $7.50 \cdot 10^{-5}
M_\odot$ (the last, 170-th zone). It was found in this case that
the equation (\ref{pex}) is quite inaccurate (evidently because
$\Delta R/R$ is not very small). As a result, the evolution time
on the presupernova stage increased nearly twice. For the variants
with non-zero $\alpha_{\rm c}$ the mass of the model reached
$1.3658 M_\odot$ after $t \simeq 7.2 \cdot 10^4 $ yr, and the
number of the Lagrangean zones became $N=170$. After that we have
added no Lagrangean zones, but simulated accretion by increasing
the outer pressure. In the runaway point (after $ t \simeq 7.9
\cdot 10^4$ yr since the beginning of calculations) the outer
pressure corresponded to the additional external mass from $2.316
\cdot 10^{-3} M_\odot$ (for $\alpha_c=3.0\cdot 10^{-4}$) to $2.342
\cdot 10^{-3} M_\odot$ (for $\alpha_c=4.0\cdot 10^{-3}$). In the
control variant with $\alpha_{\rm c}=0$ the temperature in the
central zone increased to $5\cdot 10^9 {\rm K}$ already after
$6.789 \cdot 10^4$ years, therefore the model mass could grow only
up to $1.3636 M_\odot$, which corresponds to the number of zones
$N=161$.

Let us consider the boundary condition
 (\ref{pex}) during the explosion for the variant with
$\alpha_c=1.0\cdot 10^{-3}$ when $\Delta M = 2.338\cdot 10^{-3} M_\odot$,
and $R_0 = 1.85\cdot 10^8$ cm (Fig.~1):
\begin{equation}\label{pb133m}
  P_{\rm b}(1.33 M_\odot) = P_0\left(\frac{R_0}{R}\right)^4  \; .
\end{equation}
where $P_0=5.72\cdot 10^{22}$din cm$^{-2}$. This value is to be
compared with the central pressure of the C--O core which has the
well-known lower estimate $P_{\rm c}(0) > GM^2/(8\pi R^4)$. Thus,
the ratio of the outer boundary pressure to the central one,
$P_{\rm b}(M)/P_{\rm c}(0)$ $<$ $2\Delta M/M = 3.42 \cdot
10^{-3}$, is surely very small at the beginning of the runaway. In
the subsequent runaway process it certainly changes and can
increase by factor of a few, but nevertheless remains small enough
which can be seen from the numerical results.

It is easy to estimate that the
relative contribution of the inertial term into the derived
boundary condition
 (\ref{pb133m})
with characteristic parameters of C--O core pulsations $ (4v_{\rm
p}/\tau_{\rm p})\cdot \left(R_0^2/(GM)\right)$ will be small: no
more than a few percent. Thus we can neglect it in (\ref{pex})
even when the pulsations amplitude is maximal $\Delta R \simeq R_0
$ and $\tau_{\rm p}\simeq 5{\rm c}$ by the end of deflagration
burning (See Fig.~4). This boundary condition imposed by the mass
accretion onto  the C--O core surface is in close agreement with
the outer boundary condition assumed in the previous works
(Ivanova et al., 1974; Ivanova et al., 1977a) with
 the same dependence on an outer radius $R$ :
$P_{\rm b}(M)\propto R^{-4}$ (\ref{pb133m}). Let us remind that in
the cited works the outer pressure also simulated the presence of
the outer C--O core envelope. It was taken small enough, but about
a factor of 2 higher than the value from (\ref{pb133m}). Thus, the
effect of the outer boundary was  more pronounced there than in
the present work, but less significant than in
(Dunina-Barkovskaya, Imshennik, 2000). It should be stressed that
the non-zero outer pressure $P_{\rm b}(M)\neq 0$ violates the
energy conservation law during the explosion (see Fig.~4, 5) when
the pulsations arise and the radius $R$ of the C--O core changes.
Naturally, accounting for the accreting matter in the outer
boundary condition (\ref{pex}) and (\ref{pb133m}) we imply its
attachment to the evolving C--O core without any further energy
release, i.e. merely an adhesion. In fact, this substance has,
generally speaking, another chemical composition, and can go
through the thermonuclear reactions with an energy liberation and
even with partial ejection of the accreted material. We simply
neglect those processes. On the other hand,  the role of the outer
boundary condition (\ref{pb133m}) is very significant during the
explosion, especially when the detonation wave forms near the
surface (see below).

\section*{The discussion of the results}

In the present work we have systematically  investigated
the models numerically with different values of the mixing length parameter
(\ref{lmixdri}) equal to
 $\alpha_{\rm c}=4.0\cdot 10^{-3}$; $3.0\cdot 10^{-3}$;
$2.0\cdot 10^{-3}$; $1.0\cdot 10^{-3}$;  $3.0\cdot 10^{-4}$. From
a physical standpoint, it is very difficult to choose a certain
value  $\alpha_{\rm c}$ from this wide range in the used
approximate theory of non-adiabatic convection. The energy release
in the thermonuclear reaction of carbon burning $^{12}$C$+^{12}$C
$\rightarrow ^{24}$Mg$ + \gamma$ was taken, first, by a rather
crude formula obtained by Fowler and Hoyle (1965) with electron
screening factor by Salpeter (1954) (see also Salpeter and Van
Horn 1969, and Arnett 1969), and, second, with a maximum possible
energy liberation corresponding to the instantaneous combustion of
all the carbon-oxygen mixture to $^{56}$Ni. The energy release
accepted here was taken from our early work (Ivanova et al. 1974).
It is convenient for the further comparison of our previous
results with those obtained in the current work. Such a
simplification of a thermonuclear energy generation cannot affect
essentially the obtained results due to the comparatively small
sensitivity of the ignition conditions for the C--O mixture to all
pre-exponential values in the expression for the thermonuclear
burning rate.

As a result of the presupernova evolution calculations we obtained
that just prior to the runaway the density drops to a value which
depends non-monotonically on the parameter $\alpha_{\rm c}$ and is
maximal when $\alpha_{\rm c}=4.0\cdot 10^{-3}$. In general, the
initial critical central density (in units $10^9\mbox{g cm}^{-3}$)
$\rho_{\rm c\;cr\;9}\simeq (1.88 - 2.03)$ turned out to be close
to the one accepted earlier by Ivanova et al. (1974) and later by
Ivanova et al. (1977a) ($\rho_{\rm c9}=2.33$), though it is
somewhat lower. Let us remind that in 1970s the most popular
viewpoint was that single intermediate mass stars evolve along the
convergent Pazcy\'nski's (1970) track after the formation of the
carbon (or carbon-oxygen) core. The modern approach to the
presupernova evolution involves the evolution in a close binary
system with an accreting white dwarf. In this case, according to
Yungelson (1998), all the stars in the initial main sequence mass
range $2.5\le M/M_\odot \le 10$ may become carbon-oxygen white
dwarfs in close binary systems. This range does not differ
significantly from the similar range for single stars $3.5\le
M/M_\odot \le 8$ considered previously (Pazcy\'nski, 1970).

\begin{figure}[t]
\includegraphics[scale=0.45]{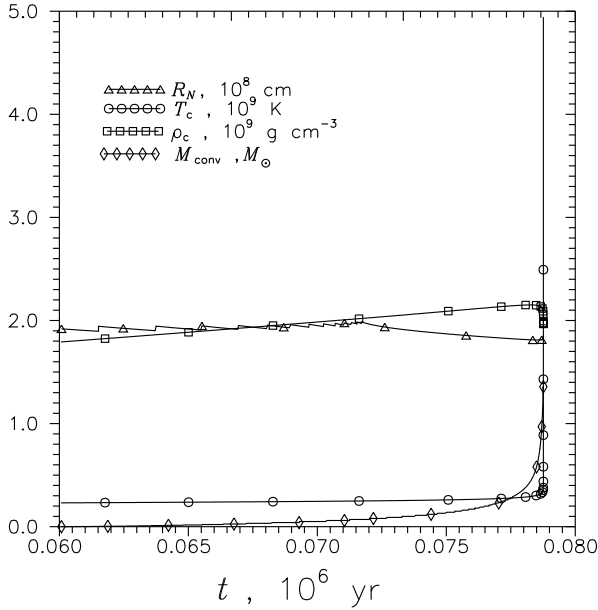}
\caption{The presupernova evolution: the dependence of the central
temperature $T_{\rm c}$, the central density $\rho_{\rm c}$, the
convective core mass $M_{\rm conv}$, and the radius $R_N$ of the
$N$-th (last) Lagrangean zone upon time for $\alpha_{\rm
c}=1.0\cdot 10^{-3}$ after the convective core formation. }
\end{figure}

Some quantities characterizing the C--O stellar core
before the runaway
are represented in Fig.~1: the convective core mass $M_{\rm conv}$,
the central temperature $T_{\rm c}$,
the central density $\rho_{\rm c}$,
and the radius of the C--O core $R_N$
over the evolution time from
$6\cdot 10^4$ yr (the beginning of the convective
core formation) to  $7.87\cdot 10^4$ yr for $\alpha_c=1.0\cdot 10^{-3}$.
Small jumps in the radius of the last ($N$-th) Lagrangean zone
shown on Fig.~2 clearly indicate the procedure
of adding of new Lagrangean zones described above.
The phase of the slow expansion before the runaway (see above)
is almost unnoticeable on this graph
because its duration ($\sim 3\cdot 10^2$ yr)
is small in comparison to the overall evolution time
(during which the star, as a whole, was contracting
very slowly) and practically does not depend on $\alpha_{\rm c}$.

\begin{figure}[t]
\includegraphics[scale=0.45]{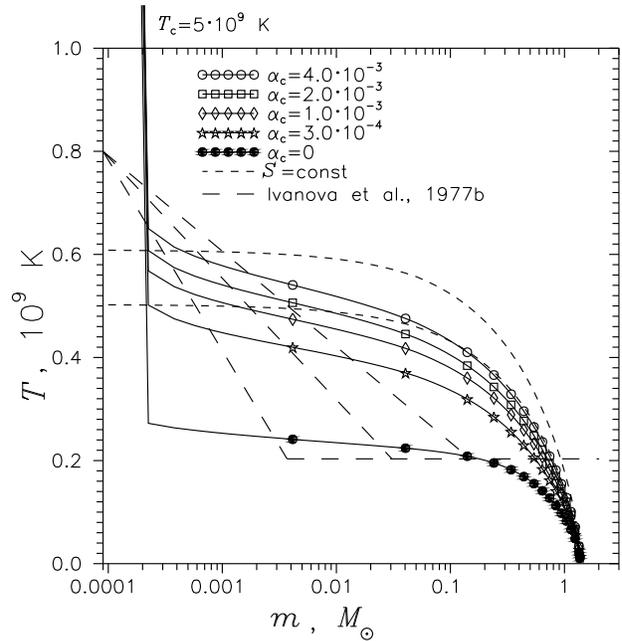}
\caption{The temperature profiles at the beginning of the SN stage
 (when the central temperature reaches $5\cdot 10^9 K$)
for various values of the parameter $\alpha_{\rm c}$.  }
\end{figure}

In Fig.~2, the initial profiles of temperature depending on the
mass coordinate are represented at the moment of the beginning of
explosion (see above). It can be seen that the temperature in the
convective variants is appreciably higher than that in the variant
without convection. Even for $\alpha_{\rm c}=3.0\cdot 10^{-4}$ the
temperature exceeds this value almost by a factor of 2 in the
central part of the convective core. Let us notice that in this
moment there is an active combustion only in the first mass zone.
On Fig.~2 we compare the initial temperature profiles with data
from Ivanova et al. (1977b). In the latter work the explosion
developed for the two (upper) $T$-profiles. Therefore, it is not
surprising that  the runaway led to the total disruption of the
star for all $\alpha_{\rm c}\ge 3.0\cdot 10^{-4}$ in the present
calculation (see below). For the variants with $\alpha_{\rm
c}=2.0\cdot 10^{-3}$ and with $\alpha_{\rm c}=3.0\cdot 10^{-4}$
which are the boundary values for the deflagration regime with
pulsations (see below), the adiabatic temperature profiles are
also shown, with the entropy equal to the entropy of the 2nd mass
zone for each variant. It can be seen that our profiles lie quite
below the corresponding adiabatic profiles, and they have another
shape for $m < 0.1 M_\odot$, which of course can affect the
development of the runaway without the regime of spontaneous
burning (Blinnikov, Khokhlov, 1986).

\begin{figure}[t]
\includegraphics[scale=0.41]{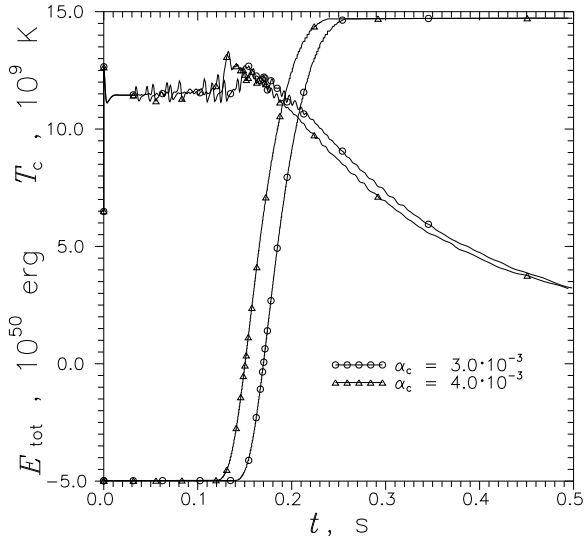}
\caption{The time dependence of the central temperature
 $T_{\rm c}$ and of the full energy of the C--O core
 $E_{\rm tot}$ for $\alpha_{\rm c}=4.0\cdot 10^{-3}$ and
$3.0 \cdot 10^{-3}$ (prompt detonation).}
 \end{figure}

It is interesting that in the second mass zone
 (with  $m=2.24\cdot 10^{-4} M_\odot$) the smooth temperature profile
for the maximum $\alpha_{\rm c}=4.0\cdot 10^{-3}$ touches the third
(lower) steep profile for which the explosion in (Ivanova et al., 1977b)
failed. But the initial profile for the variant with the minimal
$\alpha_{\rm c} = 3.0\cdot 10^{-4}$ which in fact "hardly" exploded
(see Fig.~5) already lies noticeably below the third profile from
(Ivanova et al., 1977b) for $m < 7\cdot 10^{-4} M_\odot$ (in the
first 4 mass zones). We can conclude that there is a qualitative
agreement between our present results and the calculations in the
cited work.

The development of
the runaway was different for
 $\alpha_{\rm c} \ge 3.0\cdot 10^{-3} $ (Fig.~3)
and for $\alpha_{\rm c} \le 2.0\cdot 10^{-3}$ (Fig.~4).
On the said
graphs the time dependence of the central temperature $T_{\rm c}$
and of the full energy of the C--O core $E_{\rm tot}$ on time is
shown. The latter eventually becomes equal to the value $\sim
1.5\cdot 10^{51} {\rm erg}$ corresponding to the full incineration
into $^{56}$Ni for all the variants regardless of the value
$\alpha_{\rm c}$. The value of $E_{\rm tot}$ at the moment of the
runaway beginning ($-4.3\cdot 10^{50}{\rm erg}$) is also
practically the same for all this variants.

\begin{figure}[t]
\includegraphics[scale=0.41]{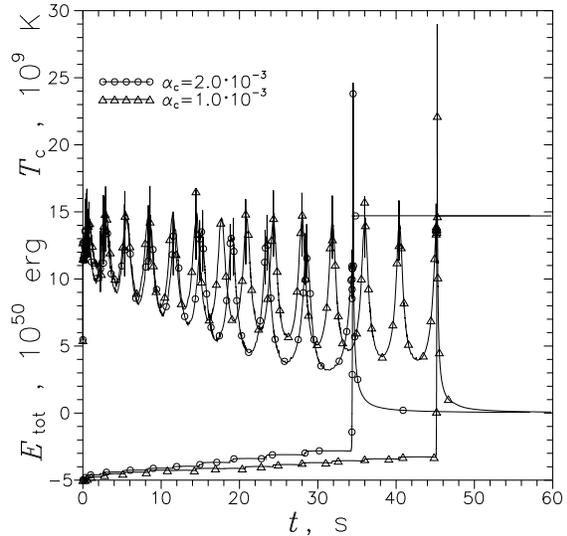}
\caption{The time dependence of the central temperature
 $T_{\rm c}$ and of the full energy of the C--O core
$E_{\rm tot}$ for $\alpha_{\rm c}=2.0 \cdot 10^{-3}$
and $\alpha_{\rm c}=1.0 \cdot 10^{-3}$
 (pulsational deflagration with delayed detonation).}
 \end{figure}

\begin{figure}[t]
\includegraphics[scale=0.4]{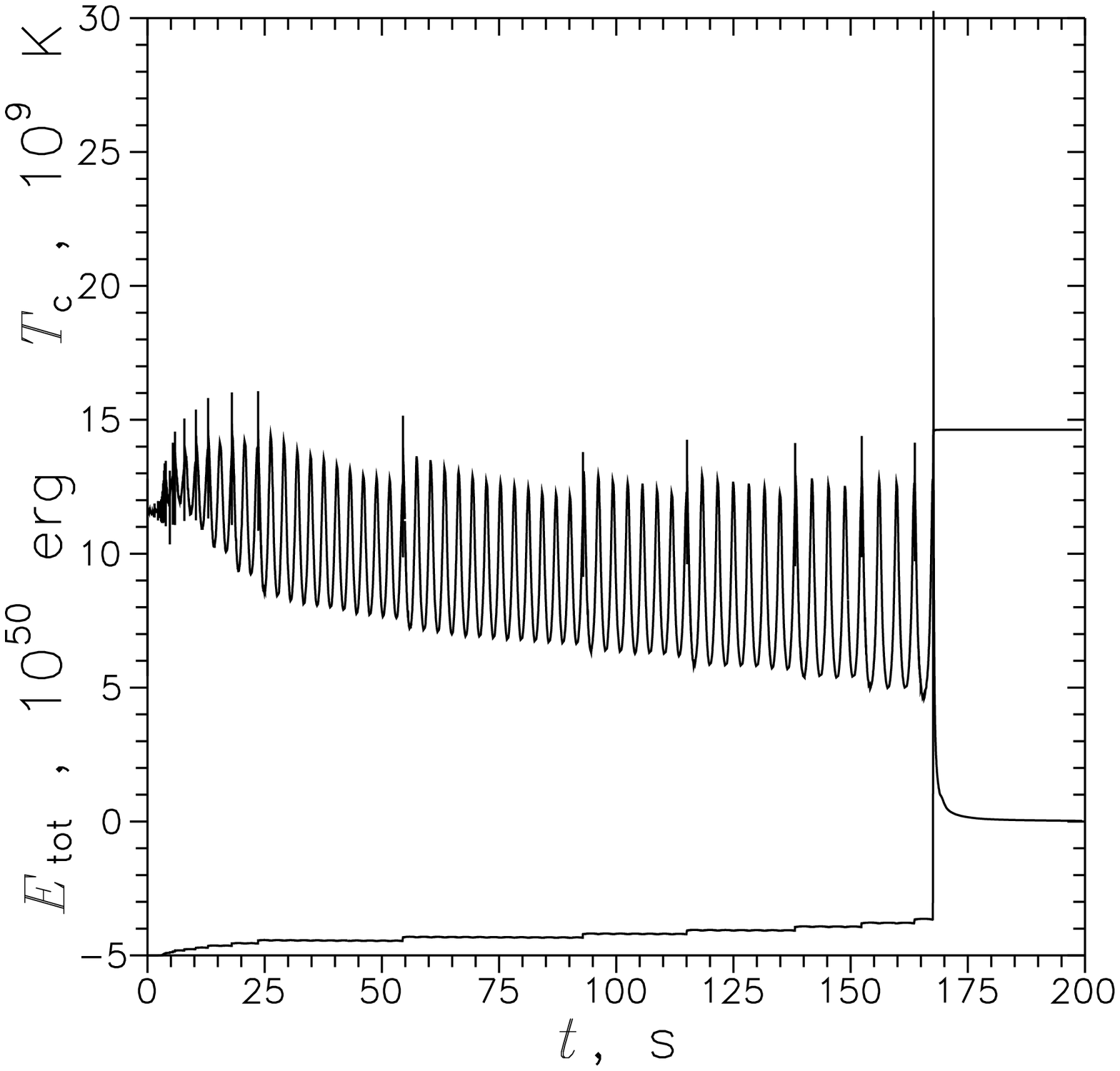}
\caption{The time dependence of the central temperature
 $T_{\rm c}$ and of the full energy of the C--O core
$E_{\rm tot}$ for $\alpha_{\rm c}=3.0 \cdot 10^{-4}$
(pulsational deflagration with delayed detonation).}
 \end{figure}

\begin{figure}[t]
\includegraphics[scale=0.4]{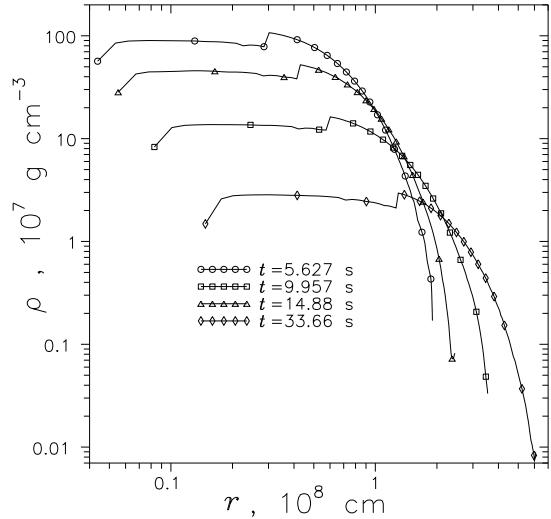}
\caption{The density profiles versus radius
 for $\alpha_{\rm c}=1.0 \cdot 10^{-3}$
 (pulsational deflagration). }
 \end{figure}

\begin{figure}[t]
\includegraphics[scale=0.4]{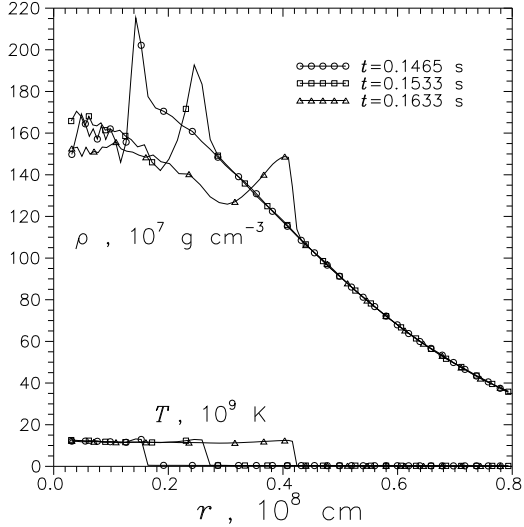}
\caption{The density and temperature profiles versus radius
 for $\alpha_{\rm c}= 3.0 \cdot 10^{-3}$
 (prompt detonation). }
 \end{figure}

\begin{figure}[t]
\includegraphics[scale=0.4]{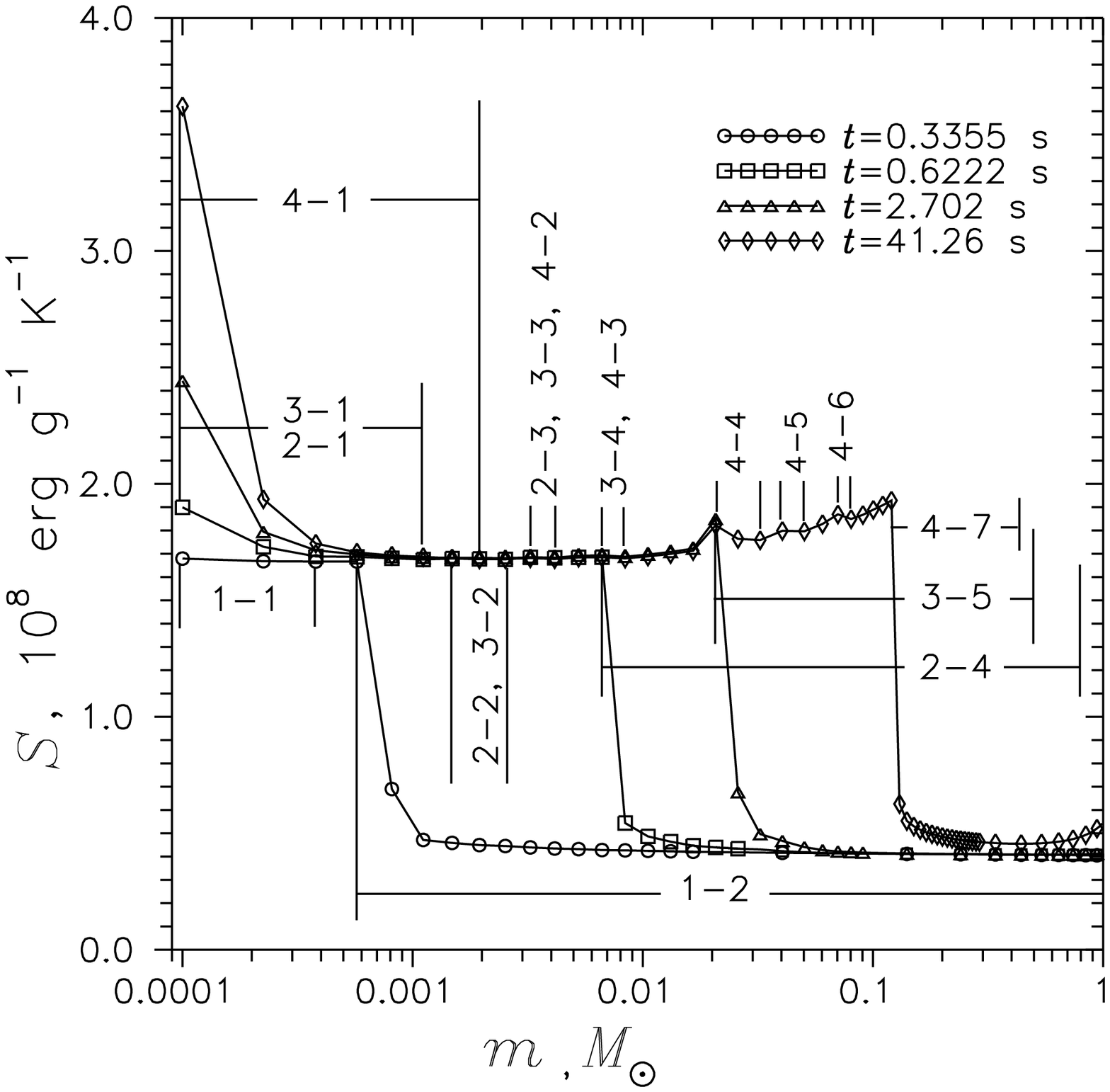}
\caption{The entropy profiles versus mass
 for $\alpha_{\rm c}=1.0 \cdot 10^{-3}$ (before the detonation).
 Convective zones are shown by labels
 and numbers, e.g. 2-1 means the first convective zone for the second
 moment (t=0.6222 s).}
 \end{figure}

The basic result of our numerical calculations is that the
pulsational deflagration regime eventuates when the mixing length
is small enough, namely $\alpha_{\rm c} \le 2.0\cdot 10^{-3}$. It
was found that for $\alpha_{\rm c} \ge 3.0\cdot 10^{-3}$ the
prompt detonation occurs (see Fig.~3). Hereby we obtained a
reasonably accurate threshold for the transition from the
detonation to the pulsational deflagration.

On Fig.~4 there is 10 pulsations for the variant with
$\alpha_{\rm c} = 2.0 \cdot 10^{-3}$
 and 14 pulsations for the variant with
$\alpha_{\rm c} = 1.0 \cdot 10^{-3}$.
In both cases the burning ends with a powerful energy release in the
last pulsation. In this case the abrupt change of the burning
regime occurs. In the previous pulsations there was a deflagration
regime, and in the last pulsation the all remaining fuel (about
90\%) is burning in the regime of detonation.
The value
$T_{\rm c}$ reaches its maximal value $(2.5 - 2.9)\cdot 10^{10}{\rm K}$
a bit later, immediately after the collision of the
detonation front propagating inward from the surface
and deflagration front,
during the afterburning in the
central zone). Thus, the hybrid
regime of the thermonuclear explosion of a carbon-oxygen white
dwarf takes place, which is most promising for the explosive
nucleosynthesis (Niemeyer and Woosley, 1997). On the other hand,
Fig.~3 demonstrates the burning without pulsations in an ordinary
detonation regime starting from the center
(for $\alpha_{\rm c}\ge 3.0\cdot 10^{-3}$).

In Fig.~5, the time dependence of the central temperature and of
the total energy is shown for the computation with $\alpha_{\rm
c}=3.0\cdot 10^{-4}$ . Here the total
number of pulsations is appreciably larger than on the previous
graph, but only 14 of them were accompanied by the propagation of
deflagration into a neighboring mass zone,
which can be well seen by very narrow specific
maxima of the temperature $T_{\rm c }$.
Let us notice that these narrow maxima correspond (due to the
energy generation) to the small jumps of the value $E_{\rm tot}$.
In the variant with
$\alpha_{\rm c}=1.0\cdot 10^{-3}$ one to three mass zones were
burning during each pulsation (except for 5th and 7th),
so almost all the pulsations had the said
temperature peaks (Fig.~4).

The deflagration regime is distinctly characterized by the
profiles of density which are represented versus  the Eulerian
radius $r$ in
Fig.~6 for $\alpha_{\rm c}=1.0\cdot 10^{-3}$.
 The density drops
 by a factor of $1.5$ at the deflagration front.
Depending upon the pulsation phase the front either moves inward
(the contraction phase) or outward (the expansion phase) during
the overall propagation (by mass zones).
The pulsations can be seen more clearly by the behaviour of
the outer layers of the star, in the still unburnt matter.
The amplitude of the outer radius pulsations grows
in time and attains the value of $\sim 1.5$
(this can be seen well due to the logarithmic scale for density
in Fig.~6).

The density profiles $\rho(r)$ for stronger convection are given in
Fig.~7 for $\alpha_{\rm c}=3.0\cdot 10^{-3}$.
The profiles of the temperature $T(r)$ are
plotted on this graph
in a small scale,
they show  neatly  the position of the burning front as a jump of $T$.
It can be seen that at the burning front the density of matter
distinctly increases
displaying the detonation burning regime.
The supersonic character of detonation is also revealed
by the quiescence of stellar layers outside the burning front.

The convective processes are most neatly characterized by the
profiles of the specific entropy which are shown in Fig.~8
for the case
$\alpha_{\rm c}=1.0\cdot 10^{-3}$
versus the mass coordinate $m$ (in logarithmic scale).
The convective zones which are defined by the known
Schwarzschild criterion, i.e. correspond to the regions with
the negative entropy gradient, are marked on these profiles
for some typical moments of time.
This figure shows that the primary convective core during
the runaway is being separated into some convective zones,
the largest of which includes one of two mass zones behind the
burning front and the rest of the convective core before the front.
The separation process is of nonstationary character and indicates
the importance of the time-dependent convection.

The entropy profiles are given in Fig.~8 before the formation of
the detonation wave propagating towards the center (see Fig.~4).
After the formation inception of such a wave, naturally, in a very
short time  $\sim 0.1$ s, the entropy behind the detonation front
rises to the values which are typical also for the deflagration,
$(2-3)\cdot 10^8 \mbox{erg g}^{-1} {\rm K}^{-1})$, i.e. by factor
of 4-6 more than an initial entropy ($\sim 0.5\cdot 10^8 \mbox{erg
g}^{-1}{\rm K}^{-1})$, with a positive gradient. Only in outer
layers (with $m\ga 1.36 M_\odot$) the negative entropy gradient
persists, and, therefore, the convection occurs, partly preventing
the detonation process  described here.

 The important role of the convection intensity characterized by a
parameter $\alpha_{\rm c}$ attracts our special
attention.
It is clear that the deflagration burning regime
with pulsations (of course,
 the deflagration is not necessarily accompanied by pulsations,
but for the latter the subsonic deflagration is the necessary
condition) takes place in the whole range of the $\alpha_{\rm c}$
values, $\alpha_{\rm c\; min} \la \alpha_{\rm c} \la 2.0\cdot
10^{-3}$. It is essential that $\alpha_{\rm c\; min}$ is above
zero, because for $\alpha_{\rm c}=0$ in our calculation there was
no hydrodynamic explosion at all, although the first mass zone was
burning. We note that for higher critical densities $\rho_{\rm c
\; cr}$ which is the case for lower accretion rate (see above),
the value of $\alpha_{\rm c\; min}$ can drop practically down to
zero (Zmitrenko et al., 1978). However, for high densities
$\rho_{\rm c \; cr}$ the account for neutrino energy losses
together with the neutronization kinetics is required (Ivanova et
al., 1977a,c). But in the present case ($\rho_{\rm c \; cr}\simeq
2\cdot 10^9\mbox{g cm}^{-3}$) the role of these complicated
processes is not important, which was shown by direct comparison
of the calculations of Ivanova et al. (1974) with those of Ivanova
et al. (1977b) at the close value of central density $\rho_{\rm
c}=2.33\cdot 10^9\mbox{g cm}^{-3}$. Thus, the account for
convection processes can prevent the pulsations development during
the deflagration burning when the parameter $\alpha_{\rm c}$
exceed some critical value $\alpha_{\rm c\; crit}$. In the present
work it is found that $\alpha_{\rm c\; crit}\ga 3.0\cdot 10^{-3}$.

\section*{A qualitative analysis of the physical conditions of
 the delayed detonation}

\begin{figure}[t]
\includegraphics[scale=0.4]{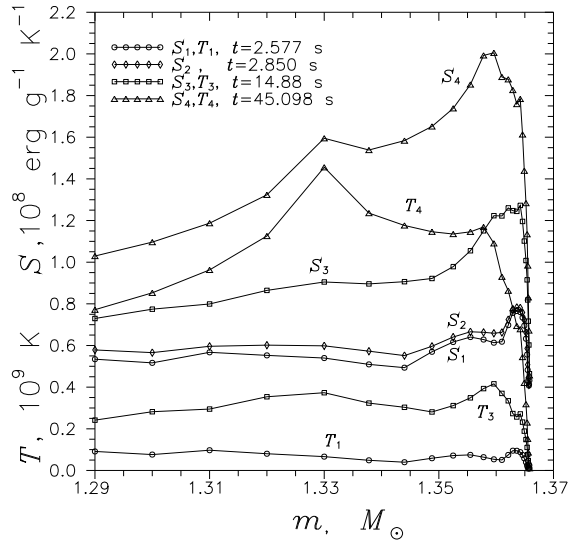}
\caption{The entropy profiles versus mass in the outer layers of the
C--O core
 for $\alpha_{\rm c}=1.0 \cdot 10^{-3}$ during the delayed detonation
(up to the beginning of last pulsation with the detonation initiation).}
  \end{figure}

\begin{figure}[t]
\includegraphics[scale=0.4]{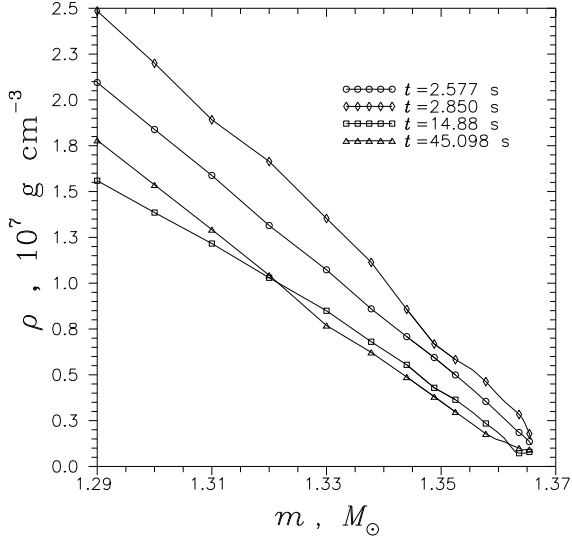}
\caption{The density profiles versus mass in the outer layers of the
C--O core
 for $\alpha_{\rm c}=1.0 \cdot 10^{-3}$
(pulsational deflagration up to the detonation initiation).}
  \end{figure}

\begin{figure}[t]
\includegraphics[scale=0.4]{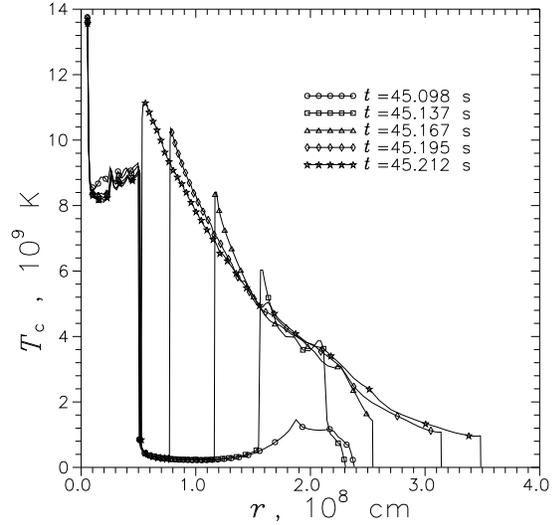}
\caption{The temperature profiles versus radius during the delayed
detonation for $\alpha_{\rm c}=1.0 \cdot 10^{-3}$.}
  \end{figure}

In essence, the emergence of the detonation, after a long period
of the deflagration burning accompanied by the development of the
global pulsations of the C--O core, can be termed a "delayed
detonation" (Niemeyer and Woosley, 1997): from the moment of the
central ignition a long time (tens of seconds) passes, i.e. many
thermodynamic time-scales of the order of the pulsation period,
2-3 s. An analysis of the results has shown that the detonation
burning front emerges in a mass coordinate about $1.33 M_\odot$,
more specifiacally, in the 150th mass zone when the total number
of zones equals 170. This coordinate is virtually independent of
the parameter $\alpha_{\rm c}$. In our model in the 150th zone the
mass grid changed from uniform to geometrically decreasing. In the
work (Dunina-Barkovskaya, Imshennik, 2000) the 127th mass zone in
which the detonation begins had the same feature, so we have to
investigate the influence of a mass grid nonuniformity on a
detonation initiation in our future work. In the present section
we shall try to give a physical justification of the entropy
growth that leads to the detonation.

After some first pulsations, i.e. long before
the generation of the burning front, the entropy of the outer
mass zones begins to increase (Fig.~9).
At the moment $t=45.098$~s (the beginning of the last pulsation) two
local entropy maximums exist, the outermost one ($ m \simeq 1.36 M_\odot$)
and the innermost one ($ m \simeq 1.33 M_\odot$). Though the entropy in the
innermost maximum is less than in the outermost one, the temperature in
the innermost maximum during the last pulsation is higher (due to the
higher density). Just at this maximum the detonation burning front
is born (see Fig.~11).
We can see the relatively  high growth
of entropy between the
close moments $t=2.577$~s and $t=2.850$~s,
between  the maximum
and the minimum of the second (!) pulsation (see Fig.~4),
which is less by amplitude
than the subsequent pulsations.

It can be seen on the graphs for density (Fig.~10)
that the moments close to the
maxima of pulsations are accompanied by a subsequent emergence
of the compression waves on the stellar
surface. Following Ivanova et al. (1982),
one can estimate the entropy growth $\Delta S$ treating
these compression waves as weak shocks
with a pressure jump $\Delta P$. The one gets
the well-known entropy jump (Landau and Lifshitz, 1954):
\begin{equation}\label{ds}
   \Delta S = \frac{1}{12T_1}\left(\frac{\partial^2V}{\partial
   P_1^2}\right)_S (\Delta P)^3.
\end{equation}
This estimate is easy to apply in the case under consideration
because we can account only for the pressure of the completely
degenerate ultrarelativistic electron gas $P_{\rm 1e} =
K\rho_1^{4/3}$, with $K=4.90\cdot
10^{14}$cm$^3$g$^{-1/3}$s$^{-2}$, in calculation of the second
adiabatic derivative in (\ref{ds}), whence it follows  that
\begin{equation}\label{d2vdp2}
  \left(\frac{\partial^2V}{\partial P_1^2}\right)_S=
  \left(\frac{\partial^2V}{\partial P_1^2}\right)_{S=0} =
  \frac{{\rm d}^2V}
  {{\rm d}P_{\rm 1e}^2}=\frac{21}{16}\frac{1}{K^2\rho^{11/3}}.
\end{equation}
Such an approximation is possible because in the zone under
consideration with $m=1.33 M_\odot$ we can take as an estimate for
the values $\rho_1$ and $T_1$ in front of the involved weak shock
their values at the moment $t=2.577$~s, namely, $\rho_1=1.073\cdot
10^7\mbox{g cm}^{-3}$, and $T_1=6.64\cdot 10^7$K. After a simple
rearrangement we get the final expression for the specific entropy
growth in (\ref{ds}) ($S$ is in units $10^8\mbox{erg g}^{-1}{\rm
K}{-1}$, see Fig.~9):
\begin{eqnarray}
  \Delta S_8 &=& 0.536\cdot
  10^6\frac{\rho_1^{1/3}}{T_1}\left(\frac{P_2}{P_1}-1\right)^3
\nonumber \\
  &=&
  1.78\left[\left(\frac{\rho_2}{\rho_1}\right)^{4/3} -
  1\right]^3.\label{ds8}
\end{eqnarray}

In the expression (\ref{ds8}) it is necessary to specify also the
amplitude of the weak shock wave. From Fig.~10 we can get
$\rho_2=1.35\cdot 10^7\mbox{g cm}^{-3}$ (for the moment
$t=2.850$~s). Then we finally get from (\ref{ds8}): $\Delta
S_8\simeq 0.085$, whereas in Fig.~9 the entropy growth between
these two moments is $\Delta S_8\simeq 0.06$. Thereby we can
reasonably explain the entropy growth obtained in calculations as
a result of a weak shock dissipation which occurs in the region of
a rather abrupt density decrease. Closer to the surface the
entropy growth subsides, according to (\ref{ds8}), because of the
density lowering towards the star edge, and inward it subsides
too, probably due to the wave amplitude reduction (see also
Fig.~10). Nevertheless, the local entropy maximum (Fig.~9) does
not disappear with time, but persists and increases furthermore by
a factor of $\sim 2.7$ before the moment of the detonation
generation. It reaches the value of $1.6\cdot 10^8\mbox{erg
g}^{-1}{\rm K}{-1}$. Let us notice that unfortunately this cannot
be seen on Fig.~8, because the outer stellar layer are shown there
in the logarithmic scale by mass.

In Fig.~11, we represent  the temperature profiles
from the moment of the burning front initiation
($t=45.098 {\rm s}$)
to its "collision" with the deflagration front at a radius
$r\simeq 5\cdot 10^7$ cm (which corresponds to the mass from the center
$0.14 M_\odot$) at the moment $t=45.212$~s.
On this graph the beginning of the expansion of matter behind
the detonation wave can be well seen: the outer radius
rises from  $2.3\cdot 10^8$ cm  to  $3.5\cdot 10^8$ cm
at the temperature $\sim 10^9{\rm K}$.
When the detonation wave propagates in the direction of increasing
density, the temperature rises to  $10^{10}{\rm K}$, partly due
to the heat capacity decline.
The compression on the front is typically
modest: only by a factor of 1.5 (see Fig.~7). The pressure jump is not
high too: about a factor of 2 till the end. It is remarkable that
the generation of the detonation front really occurs at the innermost
entropy maximum (see Fig.~9), which corresponds in Fig.~11 to
the innermost temperature maximum at the moment $t=45.098$~s.

A fair question arises: whether such a detonation front is stable
with respect to galloping  instability considered recently by
Imshennik et al. (1999). This question requires an additional
analysis, but it is conceivable that the stability here is more
natural, because the density in front of the detonation wave
increases which counteracts the decoupling of the shock wave front
from the burning zone. There is one more argument in favor of the
stability against the multidimensional disturbances, namely, the
large enough width of the burning zone which can be compared with
the C--O core radius at the densities $\rho \sim 10^7\mbox{g
cm}^{-3}$ (Imshennik, Khokhlov, 1984). This implies also that the
burning of the C--O mixture as a matter of fact does not fuse the
elements to "iron" ("Fe") in the converging detonation wave, but
is limited, for example, by such nuclides as Si etc. The problems
of nucleosynthesis in the obtained regime of delayed detonation
should be investigated in other works.

\section*{Conclusion}

In the current study, first of all, we obtained the initial conditions,
justified by evolution calculations, for the thermonuclear runaway in the
carbon-oxygen stellar core with a mass close to the Chandrasekhar
limit
% ($M_{\rm ch}=1.43 M_\odot$), %sb correct!
due to the accretion prescribed by a constant value $5\cdot
10^{-7}M_\odot\mbox{ yr}^{-1}$, which does not contradict to the
current evolutionary studies of close binary systems. In our
simulations the runaway begins at the moment defined by the
calculation itself (and the equations do not change, except for
the account for the time-dependent convection). Thus, we do not
have to solve a complicated problem of the choice of the initial
conditions, which largely affected the inception of the runaway.
In our previous work (Ivanova et al., 1974, 1977a,b,c) the initial
conditions (i.~e. the initial temperature profiles) were defined,
strictly speaking, without an adequate justification.

The present computations include a well-known approximate model of a
nonadiabatic convection, namely, the mixing length theory with
the only arbitrary parameter $\alpha_{\rm c}$ (a ratio of the mixing
length $l_{\rm mix}$ to the radial extent of the convective zone
$\Delta r_{\rm c}$). The value of this parameter
could be, in principle, extracted
from the multidimensional
approaches  developed  currently (see Lisewski et al., 2000),
i.~e. the arbitrariness
of the approximate theory could be removed. As a result of our
computations it was found that the delayed detonation regime
takes place under quite a wide range of values
$3\cdot 10^{-4} \la \alpha_{\rm c} \la 2\cdot 10^{-3}$. The lower
boundary of this interval is defined as an upper estimate.
At higher values of a parameter $\alpha_{\rm c} \ga 3\cdot 10^{-3}$
in our calculations the ordinary detonation propagating from the
center of the stellar C--O core was obtained, which is unlikely to
take place because of the instability of the burning front etc.
It could be interesting to find out whether the specified interval
intersects with the region of effective mixing length parameters
justified by a multidimensional theory of turbulence. But already
with the parametric simulations of the flame propagation velocity
we can impose
some restrictions on the parameter $\alpha_{\rm c}$.
For example, we can notice that for $\alpha_{\rm c} = 1.0\cdot
10^{-3}$ during the burning of a few first mass zones
this velocity was close to the laminar front velocity
obtained by Woosley and Timmes (1992).

In the present work, due to an application of the new hydrodynamic
code with a variable Lagrangean (difference) grid and with account
for an outer boundary pressure conditioned by accretion,
we have obtained a scenario of the development of a delayed detonation
propagating
from the surface towards the center of the star. The detonation is,
in all probability, stable with respect to the galloping  instability,
and is conditioned by a preceding stage of  the pulsational deflagration.
This may prove to be important because of the recently emerged
scepticism (Lisewski et al., 2000)
in regard to the scenario of the deflagration to detonation
transition due to the destruction of the laminar burning front by
turbulent eddies  published earlier (Khokhlov et al., 1997).

This work has been supported in part by the ISTC grant
\# 0370 and by the RFBR grant \# 00-02-17230. 
The authors are grateful to W.Hillebrandt for the helpful discussion.


\begin{thebibliography}{}


\bibitem[]{} Arnett W.D. // Astrophys. Space Sci., 1969, v.5, p.180.
\bibitem[]{} Bisnovatyi-Kogan G.S. // Physical problems of the theory
 of stellar evolution. Moscow: Nauka, 1989.(in Russian)
\bibitem[]{} Blinnikov S.I., Bartunov O.S. //
% Non-equilibrium radiative transfer in Supernova theory:
 %   Models of Linear Type II Supernovae
  Astron. Astrophys.,
  1993, v.273, p.106.
\bibitem[]{} Blinnikov S.I., Dunina-Barkovskaya N.V.//
% Cooling of white dwarfs: a method of determining their average
% mass and a constraint on neutrino properties
%  Astron.Zh. 70 (1993) 362-371. Russ
 Astronomy Reports, 1993,  v.37,  p.187.
\bibitem[]{} Blinnikov S.I., Dunina-Barkovskaya N.V. //
           MNRAS, 1994, v.266, p.289.
\bibitem[]{} Blinnikov S.I., Dunina-Barkovskaya N.V.,
           Nadyozhin D.K. // Astrophys.J.Suppl., 1996, v.106, p.171.
\bibitem[]{} Blinnikov S.I., Khokhlov A.M. //
%Pis'ma v Astron.Zh. 1986. V.12. P.318. (in Russian)
     Sov.Astron.Lett., 1986, v.12, p.131.
\bibitem[]{} Blinnikov S.I., Rudzsky M.A. //
%Pis'ma v Astron.Zh. 1984. V.10. P.363. (in Russian)
            Sov.Astron.Lett. 1984, v.10, p.152. %-155. Engl.
%\bibitem[]{} Brayton R.K., Gustavson F.G., Hachtel G.D.//
%           Proceedings of the IEEE. 1972. V.60. P.98.
\bibitem[]{} Dunina-Barkovskaya N.V., Imshennik V.S.//
 Proceedings of Lebedev Physical Institute, 2000, v.227, p.32.
\bibitem[]{} Fowler W., Hoyle F. // Nucleosynthesis in Massive Stars
           and Supernovae. 1965. Chicago, University of Chicago Press.
\bibitem[]{} Hachisu I., Kato M., Nomoto K. //
           Astrophys. J. Lett., 1996, v.470, L97.
\bibitem[]{} Haft M., Raffelt G., Weiss A.
            //Astrophys.J., 1994, v.425, p.222.
\bibitem[]{} Iben I. // Astrophys.J., 1982, v.253, p.248.
\bibitem[]{} Imshennik V.S., Kal'yanova N.L., Koldoba A.V.,
           Chechetkin V.M. // Astronomy Letters, 1999, v.25, p.206.
\bibitem[]{} Imshennik V.S., Khokhlov A.M. //
             Sov.Astrom.Lett., 1984, v.10, p.262.
%Pis'ma v Astron.Zh.
%           1984. V.10. P.631. (in Russian)
\bibitem[]{} Ivanova L.N., Imshennik V.S., Chechetkin V.M.
           // Astrophys. Space Sci., 1974, v.31, p.497.
\bibitem[]{} Ivanova L.N., Imshennik V.S., Chechetkin V.M. //
             Sov.Astron., 1977a, v.21, p.197.
%           // Astron.Zh. 1977a. V.54. P.354. (in Russian)
\bibitem[]{} Ivanova L.N., Imshennik V.S., Chechetkin V.M. //
             Sov.Astron., 1977b, v.21, p.374.
%           // Astron.Zh. 1977b. V.54. P.661. (in Russian)
\bibitem[]{} Ivanova L.N., Imshennik V.S., Chechetkin V.M. //
             Sov.Astron., 1977c, v.21, p.571.
%           // Astron.Zh. 1977c. V.54. P.1009. (in Russian)
\bibitem[]{} Ivanova L.N., Imshennik V.S., Chechetkin V.M. //
            Sov.Astron.Lett., v.8, p.8.
%           // Pis'ma v Astron.Zh. 1982. V.8. P.17 (in Russian)
\bibitem[]{} Khokhlov A.M., Oran E.S., Wheeler J.C.//
           Astrophys.J., 1997, v.478, p.678.
\bibitem[]{} Landau L.D., Lifshits E.M. // Mechanics of continua.
           Moscow: Gos.izd-vo tekhniko-teor.lit.,1954. (in Russian)
\bibitem[]{} Lisewski A.M., Hillebrandt W., Woosley S.E. //
           Astrophys.J., 2000, v.538, p.831.
\bibitem[]{} Niemeyer J.C., Woosley S.E. //
           Astrophys.J., 1997, v.475, p.740.
\bibitem[]{} Nomoto K., Sugimoto D., Neo S. //
           Astrophys. Space Sci., 1976, v.39, P.L37.
\bibitem[]{} Paczy\'nski B. // Acta Astronomica, 1970, v.20, p.47.
\bibitem[]{} Salpeter E.E. // Australian J.Phys., 1954,
           v.7, p.373.
\bibitem[]{} Salpeter E.E., Van Horn H.M.
             //Astrophys.J., 1969, v.155, p.183.
\bibitem[]{} Schinder P.J., Schramm D.N., Wiita P.J.,
           Margolis S.H., Tubbs, D.L.//
           Astrophys.J., 1987, v.313, p.531.
\bibitem[]{} Woosley S.E. // Astrophys.J., 1997, v.476, p.801.
\bibitem[]{} Woosley S.E., Timmes F.X. // Astrophys.J.,
           1992, v.396, p.649.
\bibitem[]{} Yakovlev D.G., Shalybkov D.A. //
     Itogi nauki i tekh. ser.
          Astronomia, 1988, v.38, p.191. (in Russian)
\bibitem[]{} Yungelson L.R. // Contemporary problems of stellar evolution.
           Proceedings of the international conference in Zvenigorod.
           October 13--15 1998, p.79.
\bibitem[]{} Zmitrenko N.V., Imshennik V.S., Khlopov M.Yu., Chechetkin V.M.
            // Zhurn.Exp. i Teor.Fiz., 1978, v.48, p.589. (in Russian)

\end{thebibliography}
\end{document}